\pdfoutput=1
\documentclass[aps,floatfix,superscriptaddress,twocolumn,reprint]{revtex4-1}
\usepackage{eurosym}
\usepackage{amsbsy}
\usepackage{latexsym,epsfig,graphicx}
\usepackage{dcolumn}
\usepackage{subfigure}
\usepackage{comment}
\usepackage{color}
\usepackage{bm}
\usepackage{mathrsfs}
\usepackage{amssymb}
\usepackage{amsfonts}
\usepackage{amsmath}
\usepackage{xspace}
\usepackage{epstopdf}
\usepackage{tabularx}
\usepackage{longtable}
\usepackage{mhchem}
\usepackage[colorlinks=true, letterpaper=true, pdfstartview=FitV, linkcolor=blue, citecolor=blue, urlcolor=blue]{hyperref}

\begin{document}
\title{Quench dynamics of Hopf insulators}
\author{Haiping Hu}
\affiliation{Department of Physics and Astronomy, George Mason University, Fairfax, Virginia 22030, USA}
\affiliation{Department of Physics and Astronomy, University of Pittsburgh, Pittsburgh, Pennsylvania 15260, USA}
\author{Chao Yang}
\affiliation{Division of Physics and Applied Physics, Nanyang Technological University, Singapore 637371}
\author{Erhai Zhao}
\affiliation{Department of Physics and Astronomy, George Mason University, Fairfax, Virginia 22030, USA}
\date{\today}

\begin{abstract}
Hopf insulators are exotic topological states of matter outside the standard ten-fold way classification based on discrete symmetries. Its topology is captured by an integer invariant that describes the linking structures of the Hamiltonian in the three-dimensional momentum space. In this paper, we investigate the quantum dynamics of Hopf insulators across a sudden quench and show that the quench dynamics is characterized by a $\mathbb{Z}_2$ invariant $\nu$ which reveals a rich interplay between quantum quench and static band topology. We construct the $\mathbb{Z}_2$ topological invariant using the loop unitary operator, and prove that $\nu$ relates the pre- and post-quench Hopf invariants through $\nu=(\mathcal{L}-\mathcal{L}_0)\bmod 2$. The $\mathbb{Z}_2$ nature of the dynamical invariant is in sharp contrast to the $\mathbb{Z}$ invariant for the quench dynamics of Chern insulators in two dimensions. The non-trivial dynamical topology is further attributed to the emergence of $\pi$-defects in the phase band of the loop unitary. These $\pi$-defects are generally closed curves in the momentum-time space, for example, as nodal rings carrying Hopf charge.
\end{abstract}
\maketitle
\section{Introduction}Over the past decade, topological quantum matter \cite{review1,review2} has become an active research area in condensed matter physics. A topological phase is characterized by the bulk invariant of its ground state and the corresponding gapless surface/edge states, which are usually robust against detrimental effects such as impurities. Based on the underlying discrete symmetries and dimensionality, topological states have been categorized into ten different classes, the so-called Atland-Zirnbauer ten-fold way \cite{class1,class2,class3,class4}. Recently, much attention has been paid to the explorations of topological phases outside the standard ten-fold way. A notable example is the three-dimensional (3D) Hopf insulator \cite{hopf1,DLM,DLD,hopf2,hopf3,thesis} in the absence of any symmetries, for which the ten-fold way would normally predict a trivial insulator. Besides its theoretical importance, several proposals \cite{hopfpro1,hopfpro2} have been put forward for realizing such Hopf insulators in ultracold-atom experiments. And topological links and Hopf fibration have been observed in solid-state quantum simulator of nitrogen-vacancy centers in diamond \cite{hopfexp}. 

Compared to conventional topological insulators, the Hopf insulator is characterized by an integer topological invariant---linking number $\mathcal{L}$, arising from the unique knot topology in the 3D momentum space associated with the mathematical Hopf map \cite{hopfi}. Formally, a minimal model of Hopf insulator \cite{hopf1,hopf2,DLM,DLD,hopf3,thesis} is represented by a 3D two-band Hamiltonian $H(\bm k)=\hat{\bm n}(\bm k)\cdot\bm\sigma$, where the 3D unit vector $\hat{\bm n}(\bm k)$ bridges a mapping from the 3D Brillouin zone (BZ), which is a three-torus $\textrm{T}^3$ to the target state space of two-sphere $S^2$. By ignoring the nontrivial cycle of the torus \cite{torus} and treating $\textrm{T}^3$ as three-sphere $S^3$, the nontrivial band topology is reduced to a Hopf map and characterized by homotopy group $\pi_3(S^2)=\mathbb{Z}$. The preimage of any quantum state on the Bloch sphere is a closed curve in the 3D BZ. The Hopf invariant $\mathcal{L}$ is identical to the linking number \cite{Zee,hopfbook} of the preimages of any two quantum states. 

The studies of topological phases have also been extended to out-of-equilibrium systems, e.g., systems with periodic driving \cite{fl1,fl2,fl3,fl4,fl5,fl6,fl7,fl8,fhopf,roy,wangzhong,nyao,fdqpt,fl9,fl10,fl11,fl12,fl13} or quantum quenches, which refer to sudden changes of some Hamiltonian parameters at a specific time. Hopf insulator under periodic driving has been studied in Ref. \cite{nyao} and a novel Floquet topological phase, termed Floquet Hopf insulator has been identified. For quantum quenches, the post-quench dynamics is intimately connected to the static band topology of pre- and post-quench Hamiltonians and may exhibit nontrivial dynamical phenomena \cite{qq1,qq2,qq3,qq4,qq5,haiping,qq6,qq7,qq8,qq9,qq10,qq12,qq13,qq14,quenchexp1,quenchexp2,quenchexp3,xue1,xue2,roy2,berm}. Facilitated by the great controllability of ultracold-atom experiments, the interplay between quantum dynamics and band topology can now be investigated in laboratories \cite{reviewcold1,reviewcold2,reviewcold3}. A typical example is the recent observation of dynamical vortices and spatiotemporal Hopf links \cite{qq5} for quantum quenches of 2D Chern insulators \cite{quenchexp1,quenchexp2,quenchexp3} through the time- and momentum-resolved Bloch-state tomography \cite{azi1,azi2,tomograph1,tomograph2}. 

It is then natural to investigate the quench dynamics of Hopf insulators, in particular, how to characterize its dynamical topology and relate it to the static band topology of pre- and post-quench Hamiltonians. We expect it to differ significantly from the quench of Chern insulators \cite{qq5,haiping}. Formally, the quantum quench is governed by the time evolution $|\xi(t)\rangle=U(\bm k,t)|\xi_0\rangle=e^{-iH(\bm k)t}|\xi_0\rangle$ ($\hbar=1$), with $|\xi_0\rangle$ the initial state. The post-quench state $|\xi(t)\rangle$ defines a mapping from the (3+1)-D momentum-time space (which is a four-torus $\textrm{T}^4$) to the Bloch sphere $S^2$. Such homotopy mapping is characterized by a $\mathbb{Z}_2$ invariant. We formulate this invariant using the loop unitary $U_l$ constructed from the pre- and post-quench Hamiltonians \cite{haiping} and show the dynamical topology is characterized by the homotopy group $\pi_4(\textrm{SU(2)})=\mathbb{Z}_2$. Using the phase-band representation of the loop unitary, we further prove the $\mathbb{Z}_2$ invariant $\nu$ is related to the change of static Hopf invariant across the quench, $\nu=(\mathcal{L}-\mathcal{L}_0)\bmod 2$, where $\mathcal{L}$ ($\mathcal{L}_0$) is the Hopf invariant of the post-(pre-) quench Hamiltonian. The loop unitary also provides an intuitive interpretation of the nontrivial dynamical topology as the emergence of $\pi$-defects in the phase bands. These results, as summarized in Eqs. (\ref{nupb})(\ref{relation})(\ref{hcharge}) reveal the $\mathbb{Z}_2$ invariant and underlying dynamical topological properties from different physical aspects. As the dynamical topology of quantum quenches studied in previous works \cite{qq1,qq2,qq3,qq4,qq5,haiping,qq6,qq7,qq8,qq9,qq10,qq12,qq13,qq14,quenchexp1,quenchexp2,quenchexp3,xue1,xue2} all belong to the $\mathbb{Z}$ class labeled by an integer invariant (e.g., dynamical Chern number, momentum-time Hopf invariant), our characterization constitutes a nontrivial example of $\mathbb{Z}_2$ class of quench dynamics.

The remainder of this paper is organized as follows. In Sec. \ref{sec:hi}, we briefly review the minimal model of Hopf insulators and demonstrate the Hopf links as manifestations of static band topology. In Sec. \ref{sec:z2}, we construct the homotopy invariant $\nu$ using the loop unitary $U_l$ as well as its phase-band representation and reveal its $\mathbb{Z}_2$ nature. In Sec. \ref{sec:relation}, we relate the dynamical invariant to the static pre- and post-quench Hopf invariants by using a homotopy-equivalence relation. Sec. \ref{sec:defect} discusses the $\pi$-defects in the phase bands of $U_l$, as the origin of nontrivial dynamical topology. We show the emergence of these defects by explicit examples. In Sec. \ref{sec:geometry}, we provide a geometric perspective of the dynamical topological invariant. Sec. \ref{sec:edge} discusses the dynamical behaviors of the surface states of Hopf insulators across the quantum quench. We draw conclusions in Sec. \ref{sec:summary}.

\section{Hopf insulator}\label{sec:hi}The primary model for Hopf insulator \cite{hopf1,hopf2} is the following two-band Hamiltonian
\begin{eqnarray}\label{model}
H^{p,q}=\hat{\bm n}(\bm k)\cdot \bm \sigma,
\end{eqnarray}
with $\bm\sigma=(\sigma_x,\sigma_y,\sigma_z)$ the Pauli matrices. Here $\hat{\bm n}(\bm k)=z^{\dag}\bm\sigma z$ is a normalized pseudospin vector, $|\hat{\bm n}(\bm k)|=1$. The $\textrm{CP}^1$ field $z$ is defined as
\begin{eqnarray}
z=\left(\begin{array}{cc}
z_{\uparrow}\\
z_{\downarrow}
\end{array}
\right)=\frac{1}{\sqrt{|\eta_{\uparrow}|^{2p}+|\eta_{\downarrow}|^{2q}}}\left(\begin{array}{cc}
\eta_{\uparrow}^p\\
\eta_{\downarrow}^q
\end{array}
\right),
\end{eqnarray}
where $\eta_{\uparrow,\downarrow}$ are complex numbers given by
\begin{eqnarray}
\eta_{\uparrow}&=&\sin k_x+i\sin k_y,\notag\\
\eta_{\downarrow}&=&\sin k_z+i(\cos k_x+\cos k_y+\cos k_z-M),
\end{eqnarray}
and $p,q$ are integers. $M$ is a tunable parameter to induce topological phase transitions. As any unit vector is represented by a specific point on the Bloch sphere, $\hat{\bm n}(\bm k)$ in fact defines a mapping from the 3D BZ, i.e., $\textrm{T}^3$ to the Bloch sphere $S^2$. This mapping is indexed by the Hopf invariant defined as
\begin{eqnarray}
\mathcal{L}=-\int_{\textrm{BZ}}d^3\bm k~\bm F\cdot\bm A,
\end{eqnarray}
where $\bm A$ is the Berry connection satisfying $\bm\nabla\times\bm A=\bm F$ and $\bm F=(F_x,F_y,F_z)$ is the corresponding Berry curvature
\begin{eqnarray}
F_{\mu}=\frac{1}{8\pi}\epsilon_{\mu\nu\rho}\hat{\bm n}\cdot(\partial_{\nu}\hat{\bm n}\times\partial_{\rho}\hat{\bm n}).
\end{eqnarray}
Here $\epsilon_{\mu\nu\rho}$ is the Levi-Civita symbol. $\mu,\nu,\rho$ refer to the momentum indices and summation over repeated indices is implied.

The above $\textrm{T}^3\rightarrow S^2$ mapping can be decomposed into two steps. The first step is a mapping $\textrm{T}^3\rightarrow S^3$ defined by the $\textrm{CP}^1$ field $z$. We denote the four real numbers of $z$ as $N_1=\textrm{Re}[z_{\uparrow}]$, $N_2=\textrm{Im}[z_{\uparrow}]$, $N_3=\textrm{Re}[z_{\downarrow}]$, $N_4=\textrm{Im}[z_{\downarrow}]$, which satisfy $\sum_{j=1}^4N_j^2=1$. Hence $z$ is represented by a point on the three-sphere $S^3$. The second mapping is the well-known Hopf map $S^3\rightarrow S^2$ defined by the unit pseudospin vector $\hat{\bm n}=z^{\dag}\bm\sigma z$. From this two-step mapping, the Hopf invariant can be explicitly calculated as \cite{DLM}
\begin{eqnarray}\label{hi}
\mathcal{L}=\Bigg\{\begin{array}{l l}
0,~~~~~~~~~~|M|>3,\\
pq,~~~~~~1<|M|<3,\\
-2pq,~~~~~|M|<1.\\
\end{array}
\end{eqnarray}
with the coefficients $0,1,-2$ coming from the nontrivial wrappings in the first mapping $\textrm{T}^3\rightarrow S^3$. As $(p,q)$ can take arbitrary integers, the Hopf invariant apparently can take any integer depending on the values of $p$, $q$ and $M$. As a consequence, the Hamiltonian $H^{p,q}$ constructed above realizes Hopf insulators with arbitrary linking numbers.
\begin{figure}[t]
\centering
\includegraphics[width=3.4in]{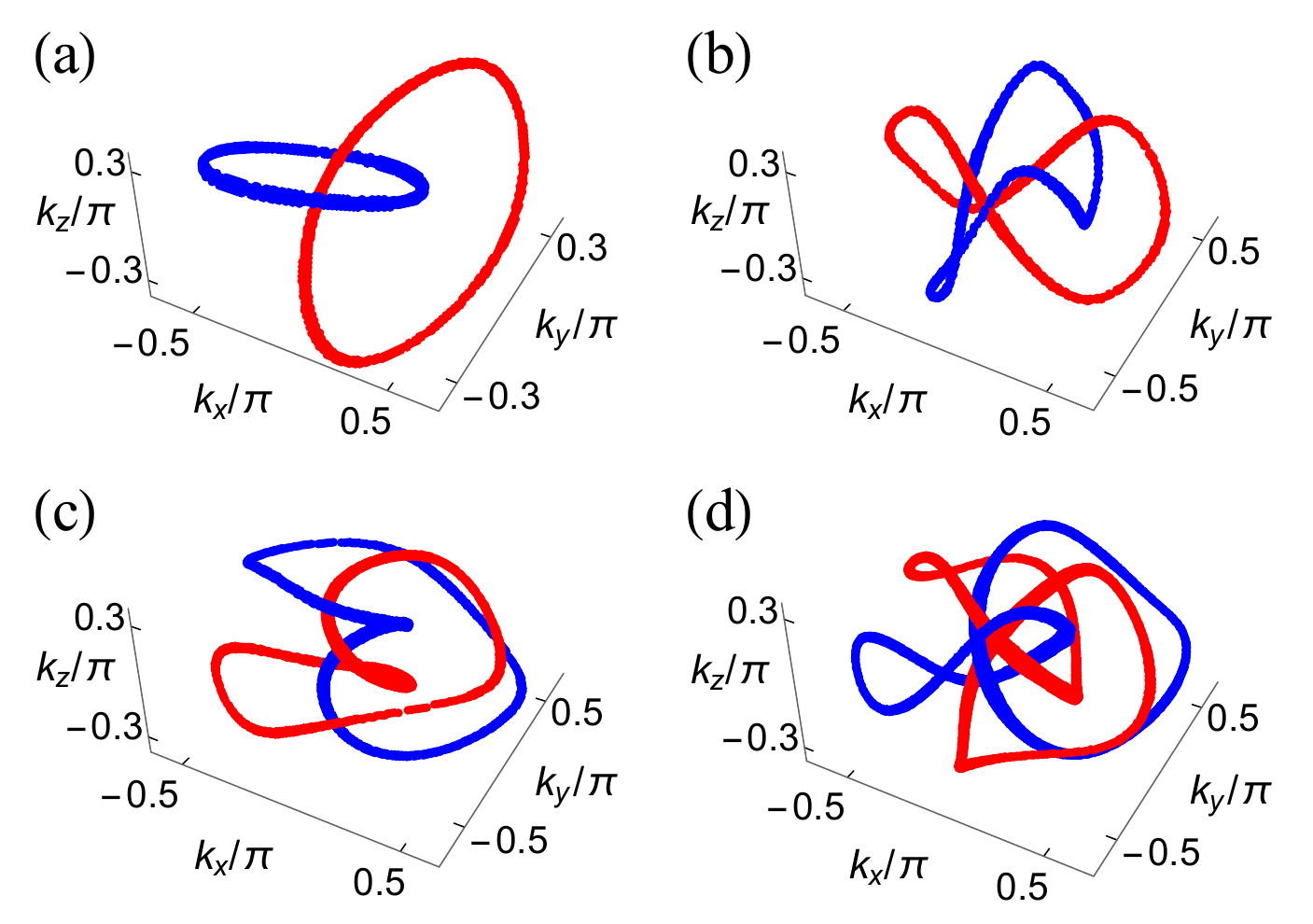}
\caption{Momentum-space Hopf links formed by the preimages of two unit Hamiltonian vectors $\hat{\bm n}=(0,\pm 1,0)$ (red and blue curves) in the 3D BZ, see model (\ref{model}). (a) $p=1,~q=1$; (b) $p=2,~q=1$; (c) $p=1,~q=2$; (d) $p=2,~q=2$. $M=2.2$ for all four cases. The linking numbers of the preimages are $\mathcal{L}=pq$.}
\label{fig1}
\end{figure}

The Hopf invariant gives the linking number of the trajectories of preimages in the 3D BZ for any two points (unit vectors) on the Bloch sphere \cite{Zee,hopfbook}. Fig. \ref{fig1} demonstrates the appearance of such Hopf links by four examples, corresponding to $(p,q)=(1,1)$, $(2,1)$, $(1,2)$, and $(2,2)$ of model (\ref{model}) with $1<M<3$. The preimages for two points $\hat{\bm n}=(0,\pm 1,0)$ on the Bloch sphere form Hopf links, with linking number $\mathcal{L}=pq$. For the $(p,q)=(1,1)$ case, the preimages link once (the simplest Hopf link) in the BZ as depicted in Fig. \ref{fig1}(a). For $(p,q)=(2,1)$, the preimages link twice and form the so-called Solomon's knot as shown in Fig. \ref{fig1}(b). For $(p,q)=(1,2)$, another Solomon's knot is formed with linking number $2$ as shown in Fig. \ref{fig1}(c). While for $(p,q)=(2,2)$, each preimage contains two components and traces two closed curves in the BZ. Each component link once with either component of the other preimage, giving rise to four linkings in total as depicted in Fig. \ref{fig1}(d). For generic $(p,q)$ and parameter $M$, the preimeges form ever more complicated Hopf links in the BZ, with linking number dictated by Eq. (\ref{hi}).

For simplicity, the model Hamiltonian Eq. (\ref{model}) is set up to have two flat bands at $E_{\bm k}=\pm 1$. More general two-band models can be considered. In describing the topology of their quench dynamics, it is convenient to carry out the following ``band flattening" procedure \cite{haiping}: we rescale the Hamiltonian $H(\bm k)\rightarrow h(\bm k)=H(\bm k)/|E_{\bm k}|$, where $\pm E_{\bm k}$ are the eigenvalues of $H(\bm k)$. 

\section{$\mathbb{Z}_2$ dynamical invariant}\label{sec:z2}
Now let us formulate the quench dynamics of Hopf insulators following the general framework established in Ref. \cite{haiping}. Without loss of generality, let us suppose the quantum quench occurs at time $t=0$ when the Hamiltonian goes through a sudden change $H_0(\bm k)\rightarrow H(\bm k)$, where $H_0(\bm k)$ and $H(\bm k)$ are called pre- and post-quench Hamiltonians, respectively. The initial state $|\xi_0\rangle$ is set as the ground state of $H_0$. We set the the post-quench Hamiltonian as the Hopf Hamiltonian $H(\bm k)=H^{p,q}$ introduced in Sec. \ref{sec:hi}. The ensuing time evolution according to the post-quench Hamiltonian is given by
\begin{eqnarray}\label{evolution}
|\xi(t)\rangle=e^{-iH(\bm k)t}|\xi_0\rangle.
\end{eqnarray}
At $t=\pi$, the time-evolved state $|\xi(t=\pi)\rangle=-|\xi_0\rangle$ returns to the initial state up to a minus sign. Thus the time evolution has period $T=\pi$ (we set $\hbar=1$ throughout). Further the two-component state $|\xi\rangle$ can be represented by a unique point on the Bloch sphere, corresponding to the unit vector
\begin{eqnarray}\label{bloch}
\bm\xi=\langle\xi(t)|\bm\sigma|\xi(t)\rangle.
\end{eqnarray}
Eqs. (\ref{evolution}) and (\ref{bloch}) together define a mapping from the (3+1)-D momentum-time space to the Bloch sphere $S^2$. The former can be regarded as a four-torus $\textrm{T}^4$ due to the periodicity of the 3D BZ and time evolution. Usually, we can ignore the nontrivial cycle of the torus and treat it as a sphere \cite{torus}. The quench dynamics is then characterized by homotopy group $\pi_4(S^2)=\mathbb{Z}_2$, yielding a reduced $\mathbb{Z}_2$ classification of the underlying dynamical topology.

Such homotopy classification does not care about the details of the base manifold, as long as it is closed and homotopic to the four-torus. The $\mathbb{Z}_2$ invariant is therefore rather general and applies to generic, gapped two-band systems. For dispersive bands, the time evolution is still periodic for each momentum $\bm k$, with a $\bm k$-dependent period $T_{\bm k}$. The base manifold $\textrm{T}^3\times [0, T_{\bm k}]$ is a deformed four-torus, the topology of which is homeomorphic to $\textrm{T}^4$. The band flattening procedure discussed above makes this explicit. By rescaling $H(\bm k)$, or equivalently rescaling time $t\rightarrow \tau= |E_{\bm k}|t$, all momentum modes have the same time period \cite{haiping}. The $(\bm k,\tau)$ space becomes $\textrm{T}^4$. While mathematically the $\mathbb{Z}_2$ invariant may be extracted from the cobordism of framed 2-manifold \cite{book}, the construction from this cobordism is complicated and abstract. Here we take another route using the recently introduced concept of loop unitary operator \cite{haiping} for generic quantum quench to construct its associated dynamical invariant. 

\subsection{Loop unitary for quantum quench}
The loop unitary $U_l$ of quantum quench is a periodized unitary operator satisfying $U_l(t=0)=U_l(t=\pi)=\textrm{I}$. $U_l$ contains both the information of the pre- and post-quench Hamiltonians. We note the time evolution operator $U(\bm k,t)$ does not have the period of the quench dynamics $T=\pi$, $U(t=\pi)\neq\textrm{I}$. Following the standard decomposition of unitary time-evolution operator into a loop part and a constant part \cite{roy}, the loop unitary is formally defined as \cite{haiping}
\begin{eqnarray}
U_l(t)=e^{-i H(\bm k) t}e^{iH_0(\bm k)t},\label{loopU}
\end{eqnarray}
where the first term on the right hand side is the time-evolution $U(\bm k,t)$ and the second term is the constant evolution according to $H_0(\bm k)$.

In contrast to the $\textrm{T}^4\rightarrow S^2$ mapping provided by the post-quench state vector $\bm\xi$, $U_l=U_l(\bm k,t)$ defines a mapping $\textrm{T}^4\rightarrow\textrm{SU(2)}$. Note that any $\textrm{SU}(2)$ matrix can be represented using the Pauli matrices as $u_0 \textrm{I}+i\bm u\cdot\bm\sigma$, with $\sum_{j=0}^3 u_j^2=1$. The $\textrm{SU}(2)$ manifold is isomorphic to the three-sphere $S^3$. The above mapping is then characterized by the homotopy group $\pi_4(\textrm{SU}(2))=\pi_4(S^3)=\mathbb{Z}_2$. An important property of the loop unitary is that $U_l(t)|\xi_0\rangle=e^{-it}|\xi(t)\rangle$, i.e., the action of loop unitary on the initial state $|\xi_0\rangle$ gives exactly the same post-quench state on the Bloch sphere with a global phase factor. We recall the homotopy relation $\pi_4(S^3)=\pi_4(S^2)$, therefore the $\mathbb{Z}_2$ invariant extracted from the homotopy group $\pi_4(\textrm{SU}(2))$ agrees with $\pi_4(S^2)$ discussed above. It can characterize the quantum dynamics and its associated dynamical topology. The loop unitary also possess other useful properties, as will be discussed later. 

A key concept related to the loop unitary is its phase band \cite{haiping,roy,fl8,phaseband,wangzhong}. The phase band is the eigenphase of the loop unitary operator. Formally for the two-band case, we can express $U_l$ in its eigenbasis as
\begin{eqnarray}\label{decom}
U_l(t)&=&e^{i\phi(\bm k,t)}|\phi_{+}\rangle\langle\phi_{+}|+e^{-i\phi(\bm k,t)}|\phi_{-}\rangle\langle\phi_{-}|.
\end{eqnarray}
Here $\pm\phi(\bm k,t)$ (with $0\leq\phi\leq\pi)$ are called phase bands and $|\phi_{\pm}\rangle$ are their corresponding eigenstates.

Generally the phase bands depend on both momentum and time and are continuous functions over the whole base manifold $\textrm{T}^4$. Two phase bands are degenerate at two special values $\phi(\bm k,t)=0,\pi$ due to the periodicity of the quasienergy zone. Such band touching points are singularities of the loop unitary \cite{phaseband} and play a crucial role for the dynamical topological properties \cite{fl8,haiping,roy,wangzhong}. For example, the phase band singularities in 3D momentum-time space resemble the Weyl points and their total charges equal the number of edge states in the corresponding quasienergy gap \cite{fl4,fl8}.

We further define the phase-band spin vector as
\begin{eqnarray}\label{svector}
\hat{\bm m}=\langle\phi_{+}|\bm\sigma|\phi_{+}\rangle.
\end{eqnarray}
In terms of $\hat{\bm m}$, the loop unitary can be rewritten as $U_l=\cos\phi+i\sin\phi~\hat{\bm m}\cdot\bm\sigma$. Comparing with Eq. (\ref{loopU}), the explicit form of phase band is determined by
\begin{eqnarray}\label{pb}
\cos\phi(\bm k,t)=\cos^2 t+\sin^2 t~\bm H(\bm k)\cdot\bm H_0(\bm k),
\end{eqnarray}
where $\bm H_0(\bm k)$ and $\bm H(\bm k)$ are pre- and post-quench Hamiltonian vectors, defined by $H_0(\bm k)=\bm H_0(\bm k)\cdot\bm\sigma$ and $H(\bm k)=\bm H(\bm k)\cdot\bm\sigma$.

\subsection{Homotopy invariant}
Now we are ready to construct the analytical form of the $\mathbb{Z}_2$ invariant to characterize the homotopy mapping $\pi_4(\textrm{SU}(2))$. The procedure follows the constructions of the dynamical invariant of Floquet Hopf insulators introduced in Ref. \cite{nyao}. This technique first embeds the $\textrm{SU}(2)$ unitary loop operator into an extended three-band unitary space $\textrm{U}(3)$ \cite{witten,embed}
\begin{eqnarray}
V(t)=\left(\begin{array}{cc}
U_l(t) & 0\\
0 & 1
\end{array}
\right).
\end{eqnarray}
The next step is to find a continuous path $\tilde{V}(t,r)$, which is parameterized by $0\leq r\leq\frac{\pi}{2}$ to contract the above three-band unitary smoothly to the trivial identity such that $\tilde{V}(t,r=0)=V(t)$ and $\tilde{V}(t,r=\frac{\pi}{2})=\textrm{I}_{3\times 3}$. The necessity of the above embedding into the extended $\textrm{U}(3)$ space is due to the triviality of homotopy mapping $\pi_4(\textrm{U}(3))$; Without the embedding, the continuous contraction of the loop unitary $U_l$ to identity $\textrm{I}_{2\times 2}$ may be forbidden due to the topological singularities hidden in the phase bands (i.e., $\pi$-defects, as will be discussed later). Remarkably, the desired $\mathbb{Z}_2$ invariant characterizing the homotopy mapping $\textrm{T}^4\rightarrow \textrm{SU}(2)$ can be computed by the following five-dimensional integral \cite{witten}
\begin{eqnarray}
\nu=\frac{-i}{240\pi^3}\int_0^{\frac{\pi}{2}}dr\int_{\textrm{T}^4} dt d^3\bm k~\textrm{Tr}(\tilde{V}^{-1}d\tilde{V})^5\bmod 2.
\end{eqnarray}
Here the integrand takes the exterior wedge product of differential forms. While the integral value may depend on the chosen contraction path, it turns out the parity of the integral is a quantized topological invariant (that is why we should take modulo $2$), independent of the specific contractions \cite{witten}.

By inserting the phase-band representation of the loop unitary and employing a band-switching contraction path \cite{nyao}, the $\mathbb{Z}_2$ invariant can be recast into
\begin{eqnarray}\label{nupb}
\nu=\frac{1}{2\pi}\int_{\textrm{T}^4}dt d^3\bm k\epsilon^{ijkl}~\partial_i\phi\mathcal{A}_j\mathcal{F}_{kl} \bmod 2,
\end{eqnarray}
where the indices $(ijkl)$ take values in $(t,k_x,k_y,k_z)$, $\mathcal{A}_j=\frac{-i}{4\pi}[\langle\phi_{+}|\partial_j\phi_{+}\rangle-\langle\partial_j\phi_{+}|\phi_+\rangle]$ is the Berry connection of the phase band, and $\mathcal{F}_{kl}=\frac{-i}{2\pi}[\langle\partial_k\phi_+|\partial_l\phi_+\rangle-\langle\partial_l\phi_+|\partial_k\phi_+\rangle]$ is the corresponding Berry curvature. The ($\bm k,t$)-dependence of the phase band $\{\phi,|\phi_{+}\rangle\}$ is explicitly given by Eqs. (\ref{loopU})(\ref{decom})(\ref{pb}).

\section{Relation with static Hopf invariant}\label{sec:relation}
Having established the $\mathbb{Z}_2$ invariant, the next question is how to relate it to the integer Hopf invariants of the pre- and post-quench Hamiltonians. In our system, however, the quench dynamics and static band topology are characterized by different types of indices. To proceed, we utilize an important property of the loop unitary \cite{haiping}: $U_l$ is homotopic \cite{roy} to the following two-step evolution generated by $H(\bm k)$ and $H_0(\bm k)$,
\begin{eqnarray}\label{cloop}
U_g =\Big\{\begin{array}{l l} e^{-2 i H(\bm k) t},~~~~~~~~~~~~\quad 0<t<\frac{\pi}{2}; \\ -e^{2i H_0(\bm k)(t-\frac{\pi}{2})},~~~~~ \quad \frac{\pi}{2}<t<\pi.\\ \end{array}
\end{eqnarray}
Note that $U_g$ is also a periodized unitary operator satisfying $U_g(t=0)=U_g(t=\pi)=\textrm{I}_{2\times 2}$. Such homotopy equivalence indicates that we can find a continuous path to connect the two loop unitaries $U_l$ and $U_g$ by preserving the relevant $\pi$-gap in the phase bands. Consequently $U_l$ and $U_g$ have the same global properties and topological invariant $\nu=\nu[U_g]$. This statement reveals the connections between quantum quench and Floquet driving systems. For Floquet Hopf insulators \cite{nyao}, the time evolution is in general not time-periodic. However the time evolution operator can always be decomposed into the product of a micro-motion operator and constant evolution operator \cite{roy}. Similar to the loop unitary, the micro-motion operator constructed for a given gap defines a $\textrm{T}^4\rightarrow \textrm{SU(2)}$ mapping. A full description of the Floquet topology of a periodically driven Hopf insulator requires two dynamical invariants extracted from the micro-motion operators, one for each gap, and an additional Hopf invariant for the quasienergy bands. The dynamical invariant for each gap dictates the appearance of Floquet surface modes.

From the above homotopy equivalence, the $\mathbb{Z}_2$ dynamical invariant can be simply computed using the phase band of $U_g$. In each time step, the phase band of $U_g$ is only linearly dependent on $\bm t$ (without any $\bm k$-dependence due to its normalized form). For $0<t<\frac{\pi}{2}$, $\phi=2t$; for $\frac{\pi}{2}<t<\pi$, $\phi=2\pi-2t$. The integral in Eq. (\ref{nupb}) then yields
\begin{eqnarray}\label{relation}
\nu&=&\frac{1}{2\pi}\bigg[\int_0^{\pi} d\phi\int_{\textrm{BZ}} d\bm k^3\epsilon^{jkl}(\mathcal{A}_j\mathcal{F}_{kl})|_{H(\bm k)}\notag\\
&&+\int_{\pi}^{0} d\phi\int_{\textrm{BZ}} d\bm k^3\epsilon^{jkl}(\mathcal{A}_j\mathcal{F}_{kl})|_{H_0(\bm k)}\bigg]\bmod 2\notag\\
&=&(\mathcal{L}-\mathcal{L}_0) \bmod 2,
\end{eqnarray}
with $\mathcal{L}$ and $\mathcal{L}_0$ the Hopf invariants of the post- and pre-quench Hamiltonians, respectively. 

Eq. (\ref{relation}) is a remarkable result. It indicates that the quench dynamics of a Hopf insulator can be categorized into two topologically distinct classes. The trivial class corresponds to when the difference between the static pre- and post- quench Hopf invairants is an even number and thus identical to zero; while for the nontrivial cases, their difference is an odd number. Here we provide a more physical perspective of such $\mathbb{Z}_2$ nature, which does not resort to homotopy group theory. To this end, consider the following continuous unitary transformation of the loop unitary $U_l$,
\begin{eqnarray}
U'_{l}(\lambda)=S^{\dag}(\lambda)U_l S(\lambda).
\end{eqnarray}
Here $S(\lambda)=e^{-i\lambda H_0(\bm k)t}$ is parameterized by $\lambda\in[0,1]$. It is clear at $\lambda=0$, $U'_l(\lambda=0)=U_l$ and at $\lambda=1$, $U'_l(\lambda=1)=e^{i H_0(\bm k)t}e^{-iH(\bm k)t}$. This unitary transformation interpolates between the above two unitaries while preserving the phase-band gap (the eigenphase is invariant under the transformation). Hence the dynamical invariant remains unchanged for any $\lambda\in[0,1]$. Using the same homotopy-equivalence argument above, we find the dynamical invariant for $U'_l(\lambda=1)$ is $-\nu$. To identify $\nu$ with $-\nu$, it must be a $\mathbb{Z}_2$ number, and modulo $2$ is necessary in Eq. (\ref{relation}). 

\section{Topological defect of the loop unitary}\label{sec:defect}
In this section, we investigate the topological defects in the loop unitary $U_l$. We first attribute the nontrivial dynamical topology to these topological singularities \cite{haiping}. By explicit examples, we show the defects are generally closed loops in the (3+1)-D momentum-time space. Further, we illustrate the emergence of defects accompanied with the topological phase transitions of the post-quench Hamiltonian.
  
\subsection{$\pi$-defect and Hopf charge}
In this subsection, we discuss the topological defects of the loop unitary $U_l$ associated with the nontrivial dynamical topology of the quench dynamics. The analysis follows the framework of Ref. \cite{haiping}. These defects are phase-band singularities located in the (3+1)-D momentum-time space and correspond to the band degeneracies. Due to the $2\pi$-periodicity of the eigenphase, there exist two types of band degeneracies \cite{fl8,roy,wangzhong,haiping,phaseband} with $\phi=-\phi$. One kind is located at $\phi=0$, dubbed zero-defect, where the two phase bands touch at the zero gap; the other kind is at $\phi=\pi$, dubbed $\pi$-defect, where the two phase bands touch at the $\pi$-gap. 

According to the discussions on homotopy equivalence in Sec. \ref{sec:relation}, the change of static Hopf invariants should be directly related to the $\pi$-defects. We focus on the $\pi$-defects from now on. It is clear from Eq. (\ref{pb}) that the $\pi$-defects are determined by the following two conditions:
\begin{eqnarray}\label{pidefect}
t=\frac{\pi}{2},~~~\bm H(\bm k)=-\bm H_0(\bm k).
\end{eqnarray}
The second condition gives the momentum points in the BZ where pre- and post-quench Hamiltonian vectors are anti-parallel to each other. As the time evolution of state after quench can be regarded as the precession of the spin vector $\bm\xi$ with respect to the post-quench Hamiltonian vector $\bm H(\bm k)$, it is obvious that at these special momenta the spin vector $\bm\xi$ stays fixed at its initial direction. 

The second condition in Eq. (\ref{pidefect}) generally determines a 1D closed curve in the BZ. The $\pi$-defect then corresponds to a loop defect in the (3+1)-D momentum-time space and for each defect, we can define its associated Hopf charge \cite{nyao}. To this end, we rewrite the topological invariant using a coordinate transformation $(t,k_x,k_y,k_z)\rightarrow (\phi, x_1,x_2,x_3)$. In this notation, the eigenphase $\phi=\phi(\bm k,t)$ depends continuously on momentum and time. The three coordinates $(x_1,x_2,x_3)$ locally label the 3D submanifold with constant $\phi$, i.e., equal-$\phi$ surface. In general this coordinate transformation is not one-to-one on the entire space $\textrm{T}^4$. If this is the case, the equal-$\phi$ surface contains multiple disjoint pieces (components) $P$, each labeled by its own local coordinates $(x_{1P},x_{2P},x_{3P})$. The final results can be obtained by summing over all different components. Under this transformation, the $\mathbb{Z}_2$ invariant in Eq. (\ref{nupb}) is recast into
\begin{eqnarray}\label{hcharge}
\nu&=&\frac{1}{2\pi}\sum_{P}\int_0^{\pi} d\phi\int d\bm x_P^3\epsilon^{jkl}\mathcal{A}_j\mathcal{F}_{kl} \bmod 2\notag\\
&=&\sum_P h_P \bmod 2.
\end{eqnarray}
In the first line, the second integral is performed for each equal-$\phi$ surface and the first integral is ramping up the corresponding eigenphase. The summation $\sum_P$ is for all components as discussed above. The second integral is nothing but the Hopf invariant over the 3D submanifold. It is independent of $\phi$ hence defines a local charge of the $\pi$-defect as $h_P=\frac{1}{2}\int d\bm x_P^3\epsilon^{jkl}\mathcal{A}_j\mathcal{F}_{kl}$. In general, the Hopf charge should be determined numerically and the 3D integral can be performed for an arbitrary 3D submanifold (not necessarily the equal-$\phi$ surface) enclosing these defects. Eq. (\ref{hcharge}) indicates that the $\mathbb{Z}_2$ dynamical invariant is nothing but the total Hopf charges of these $\pi$-defects modulo $2$. By a careful analysis of the properties of these $\pi$-defects, we can fully extract the dynamical topology of the quantum quench.

\subsection{Examples}
Now we demonstrate the emergence of $\pi$-defects in the phase bands by simple examples. For convenience, let us suppose the initial state is fully polarized $|\xi_0\rangle=(1,0)^T$ and the pre-quench Hamiltonian is $H_0=-\sigma_z$. First we consider a post-quench Hamiltonian $H(\bm k)=H^{1,1}$ as defined in Eq. (\ref{model}) with $1<M<3$. Explicitly, $H(\bm k)=\hat{\bm n}(\bm k)\cdot\bm\sigma$ with $\hat{\bm n}(\bm k)=(n_x,n_y,n_z)$ given by
\begin{eqnarray}
n_x=\frac{2}{N_0}[\sin k_x\sin k_z+\sin k_y(\sum_{\alpha=1}^3\cos k_{\alpha}-M)],\notag
\end{eqnarray}
\begin{eqnarray}
n_y=\frac{2}{N_0}[-\sin k_y\sin k_z+\sin k_x(\sum_{\alpha=1}^3\cos k_{\alpha}-M)],\notag
\end{eqnarray}
\begin{eqnarray}
n_z=\frac{1}{N_0}[\sin^2 k_x+\sin^2 k_y-\sin^2 k_z-(\sum_{\alpha=1}^3\cos k_{\alpha}-M)^2].\notag\\
\end{eqnarray}
Here $N_0=|\eta_{\uparrow}|^2+|\eta_{\downarrow}^2|$ is the normalization coefficient. The static Hopf invariants for the post- and pre- Hamiltonians are $\mathcal{L}=1$ and $\mathcal{L}_0=0$, respectively. 

\begin{figure}[t]
\centering
\includegraphics[width=3.4in]{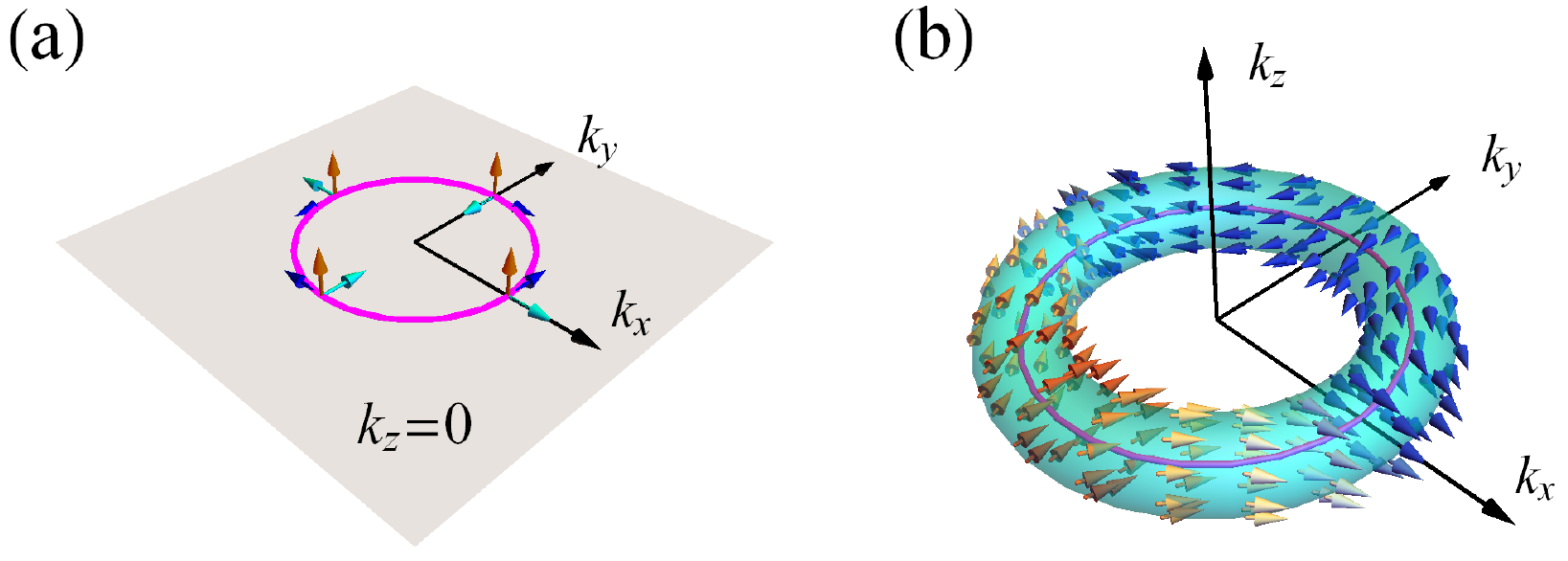}
\caption{(a) Schematics of the $\pi$-defect and Weyl-cone rotation. The $\pi$-defect forms a loop $C$ [Eq. (\ref{curve})] in the $k_x-k_y$ plane (magenta curve) with $k_z=0$ and $t=\frac{\pi}{2}$. The three arrows of different colors label the orthogonal frame defined by Eq. (\ref{twist}). (b) Plot of phase-band spin vector $\hat{\bm m}$ [defined in Eq. (\ref{svector})] on a cylinder enclosing the loop defect.}
\label{fig2}
\end{figure}

The $\pi$-defect is determined by $t=\frac{\pi}{2}$ and $\hat{\bm n}(\bm k)=(0,0,1)$. These two conditions yield a closed curve (denoted as $C$) in the $k_x-k_y$ plane:
\begin{eqnarray}\label{curve}
C:~~~t=\frac{\pi}{2},~k_z=0,~\cos k_x+\cos k_y=M-1.
\end{eqnarray}
This is a loop defect in the (3+1)-D momentum-time space. We note the $\pi$-defects in the quench dynamics of Chern insulators take the form Weyl-like degeneracy points in the (2+1)-D momentum-time space \cite{haiping}. To investigate the dispersions near this loop defect, consider $M=3-\epsilon$ ($\epsilon$ is a small quantity). The defect curve $C$ is a circle in the $k_x-k_y$ plane given by [See Fig. \ref{fig2}(a)]
\begin{eqnarray}
k_x^2+k_y^2=2\epsilon.
\end{eqnarray}
The circle is parameterized by polar angle $\theta\in[0,2\pi]$, $k_x=r\cos\theta$, $k_y=r\sin\theta$, with $r=\sqrt{2\epsilon}$. We focus on a specific point $\theta_0$ on this curve. For the points nearby, the post-quench Hamiltonian $H(\delta r,\delta k_z)=\delta \bm n\cdot\bm\sigma$ reads (to the linear order of derivations $\delta r$, $\delta k_z$),
\begin{eqnarray}
\delta n_x&=&\frac{2\cos\theta_0}{r}\delta k_z-2\sin\theta_0\delta r,\notag\\
\delta n_y&=& -\frac{2\sin\theta_0}{r}\delta k_z-2\cos\theta_0\delta r,\notag\\
\delta n_z&=&1+\textrm{O}(\delta k_z^2,\delta r^2).
\end{eqnarray}
The loop unitary near this defect takes the form (to the linear order of $\delta r$, $\delta k_z$ and $\delta t$)
\begin{eqnarray}
U(\delta r,\delta k_z,\delta t)&=&-\textrm{I}-i (\delta n_y\sigma_x-\delta n_x\sigma_y-2\delta t\sigma_z)\notag\\
&\equiv&-\textrm{I}+i\delta\bm m\cdot\bm\sigma.
\end{eqnarray}
It is clear that along the defect curve, the dispersion relations are linear along the other three orthogonal directions of $\delta r$, $\delta k_z$ and $\delta t$. Each point on this loop defect behaves like a Weyl-like degeneracy point, which is a general feature of the Hopf charge \cite{nyao}. 

Now let us travel along this defect curve. During the round trip, the Weyl cone will rotate before returning to itself. In fact, at each point along the loop, the three orthogonal vectors $\frac{\delta\bm m}{\delta r}$, $\frac{\delta\bm m}{\delta k_z}$, and $\frac{\delta\bm m}{\delta t}$ define a local orthogonal frame of the Weyl cone extending in the momentum-time space. They are
\begin{eqnarray}\label{twist}\label{twist}
\frac{1}{r}\frac{\delta\bm m}{\delta r}&=&\frac{2}{r}(\cos\theta,-\sin\theta,0),\notag\\
\frac{\delta\bm m}{\delta k_z}&=&\frac{2}{r}(\sin\theta,\cos\theta,0),\notag\\
\frac{\delta\bm m}{\delta t}&=&(0,0,2).
\end{eqnarray}
These three vectors are linearly independent and the frame twists by $2\pi$ about the time axis after a round trip along $C$, as illustrated in Fig. \ref{fig2}(a). In Fig. \ref{fig2}(b), we have plotted the phase-band spin vector $\hat{\bm m}$ on a cylinder surface enclosing the loop defect. These vectors give the direction of the effective magnetic field. The $2\pi$ rotation along the loop defect can be clearly observed.  

Above we have considered the case $M\lesssim 3$ for simplicity, where the defect loop is a circle. However the Weyl cone nature of the $\pi$-defect and the $2\pi$-twisting along the loop are valid for generic $M$. Similar calculations apply to the quantum quench with other post-quench Hamiltonians, e.g., $H(\bm k)=H^{p,1}$ ($p>1$) and the resulting $\pi$-defect is a closed curve composed of higher-order Weyl-like band degeneracies. The frame twisting along the loop defect is given in Eq. (\ref{twist}) by replacing $\theta\rightarrow p\theta$.

\begin{figure}[t]
\centering
\includegraphics[width=3.4in]{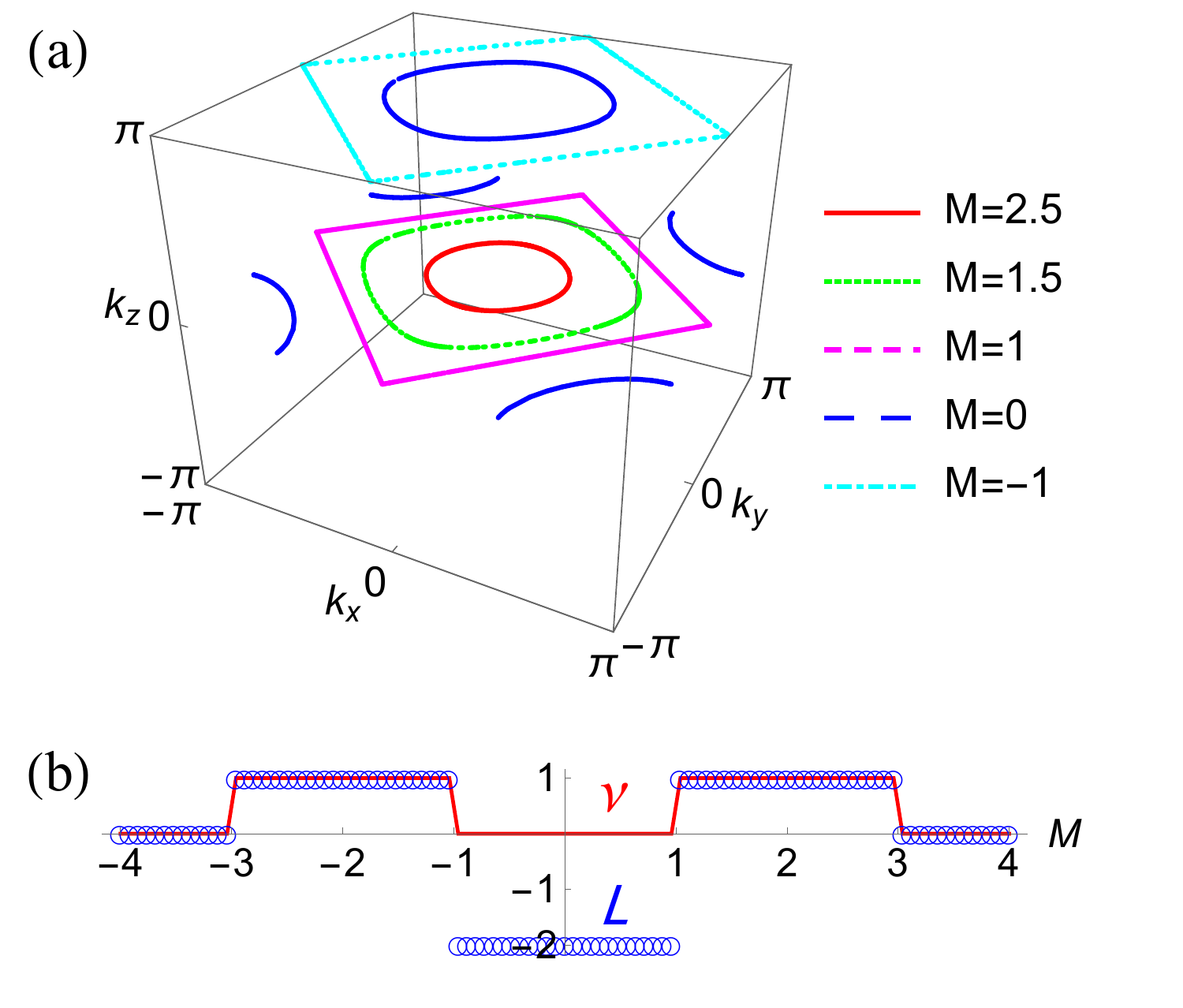}
\caption{(a) The evolution of the $\pi$-defects in the BZ with respect to $M$. For $1<M<3$, there exists only one $\pi$-defect loop centered at $\bm k=(0,0,0)$; while for $-1<M<1$, there exist two $\pi$-defect loops centered at $\bm k=(0,0,\pi)$ and $\bm k=(\pi,\pi,0)$, respectively. (b) Phase diagram with respect to $M$. The blue circle and red line represent the Hopf invariant $\mathcal{L}$ of the post-quench Hamiltonian and the dynamical invariant $\nu$, respectively. $p=q=1$.}
\label{fig3}
\end{figure}

\subsection{Topological phase transitions}
The $\pi$-defects encode the nontrivial topological properties of quantum quench. Here we demonstrate the evolution of $\pi$-defects, accompanied with the topological phase transitions of the post-quench Hamiltonian, as the tuning parameter $M$ in model (\ref{model}) is varied from $M>3$ to $M<-3$. The main results are summarized in Fig. \ref{fig3}. For $1<M<3$, there exists only one loop $\pi$-defect centered at $(k_x,k_y,k_z)=(0,0,0)$. By decreasing $M$, the loop defect expands. At $M=1$, a topological transition happens, with the static Hopf invariant $\mathcal{L}$ changing from $1$ to $-2$ according to Eq. (\ref{hi}) [see the phase diagram in Fig. \ref{fig3}(b)]. Correspondingly, the $\pi$-defect touches the boundary of the BZ. After the transition within the range $-1<M<1$, another $\pi$-defect loop emerges at the plane of $k_z=\pi$. The two loop defects are now centered at $(k_x,k_y,k_z)=(\pi,\pi,0)$ and $(0,0,\pi)$, respectively. For each $\pi$-defect, the dispersion nearby is linear, Weyl-like, giving rise to Hopf charge $1$. The total charge is $0$ due to its $\mathbb{Z}_2$ nature. With further decreasing of $M$, the defect loop at $k_z=0$ shrinks and the $\pi$-defect at $k_z=\pi$ expands. Finally at $M=-1$, the $\pi$-defect at $k_z=\pi$ touches the boundary of the BZ; while the loop at $k_z=0$ vanishes. The dynamical invariant returns to $\nu=1$ for $-3<M<-1$.  

\section{Geometric visualization of the $\mathbb{Z}_2$ invariant}\label{sec:geometry} 
In Sec. \ref{sec:relation}, we have provided an explanation of the $\mathbb{Z}_2$ nature of the quench dynamics based on the gap-preserving unitary transformation besides the mathematical homotopy mapping. Here we briefly discuss a complementary geometric perspective \cite{nyao,book} of the dynamical invariant from the properties of Jacobian twisting along preimage loops. This is possible because, both the static Hopf invariant $\mathcal{L}$ and dynamical invariant $\nu$ are homotopy invariants which can be reduced to lower dimensions. For the Hopf invariant, we have the homotopy relation:
\begin{eqnarray}
\pi_3(S^2)=\pi_1(\textrm{SO(2)})=\mathbb{Z}.
\end{eqnarray}
To see the above equivalence, we consider two infinitesimally close points $\bm n$ and $\bm n+\delta \bm n$ on the Bloch sphere, with their preimages forming closed curves parameterized by $\theta$ as $\bm k(\theta)$ and $\bm k(\theta)+\delta \bm k(\theta)$, respectively. The Hopf invariant $\mathcal{L}$ gives the linking number of these two preimages. The difference between the two preimages defines a $2\times 3$ Jacobian matrix as $J_{ij}(\theta)=\frac{\partial n_j}{\partial k_i}$, with $J(\theta)\delta \bm k(\theta)=\delta\bm n$. Note that $\bm n$ is a unit vector, only the derivatives along the two non-radial directions are nonzero. The Jacobian has two independent column vectors. By choosing one of the them to point along the direction to the second preimage (it will trace out the second preimage with the evolution of $\theta$), the twisting of the two column vectors along the first preimage then directly gives the linking number, i.e., Hopf invariant. The twisting is classified by homotopy group $\pi_1(\textrm{SO}(2))=\mathbb{Z}$.
\begin{figure}[t]
\centering
\includegraphics[width=3.4in]{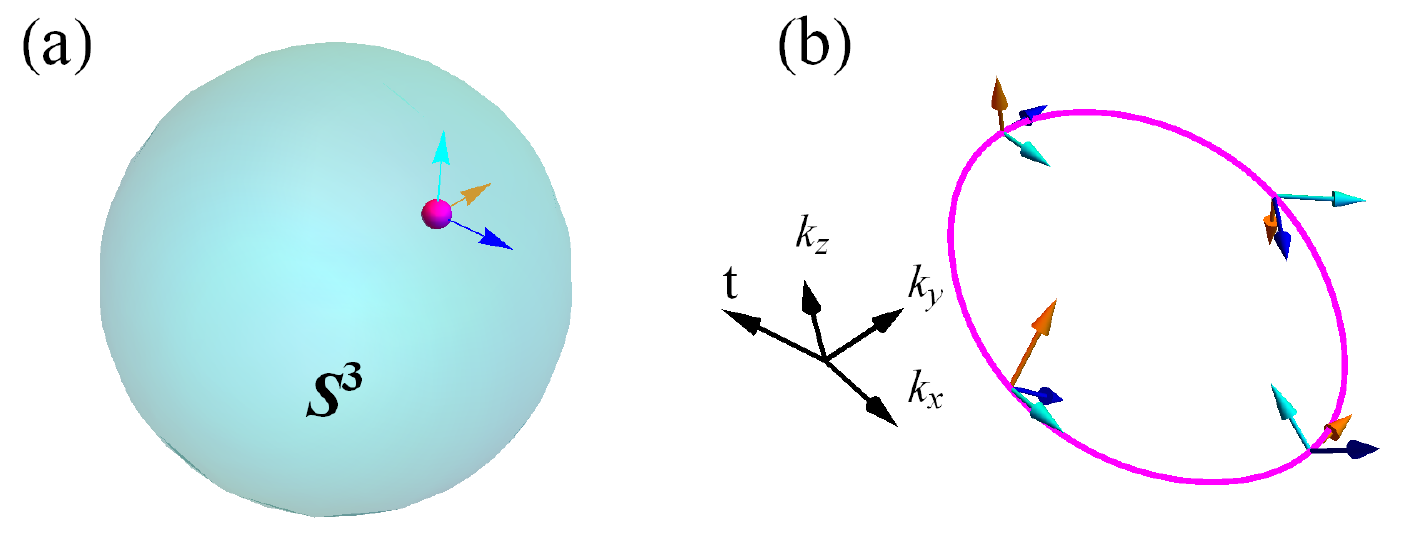}
\caption{(a) Sketch of the target manifold $\textrm{SU}(2)\cong S^3$ in the homotopy mapping $\textrm{T}^4\rightarrow\textrm{SU}(2)$ provided by the loop unitary. For an arbitrary point on $S^3$ (magenta dot), the three colored arrows label its nearby points. (b) The preimage of the chosen point, which is a closed curve (magenta) in the base manifold $\textrm{T}^4$. The end points of the colored arrows form the preimages of the three nearly points in (a). The twisting of the orthogonal frame by traveling along the preimage loop is characterized by the $\mathbb{Z}_2$ invariant $\nu$.}
\label{fig4}
\end{figure}
\begin{figure*}[!t]
\centering
\includegraphics[width=5.2in]{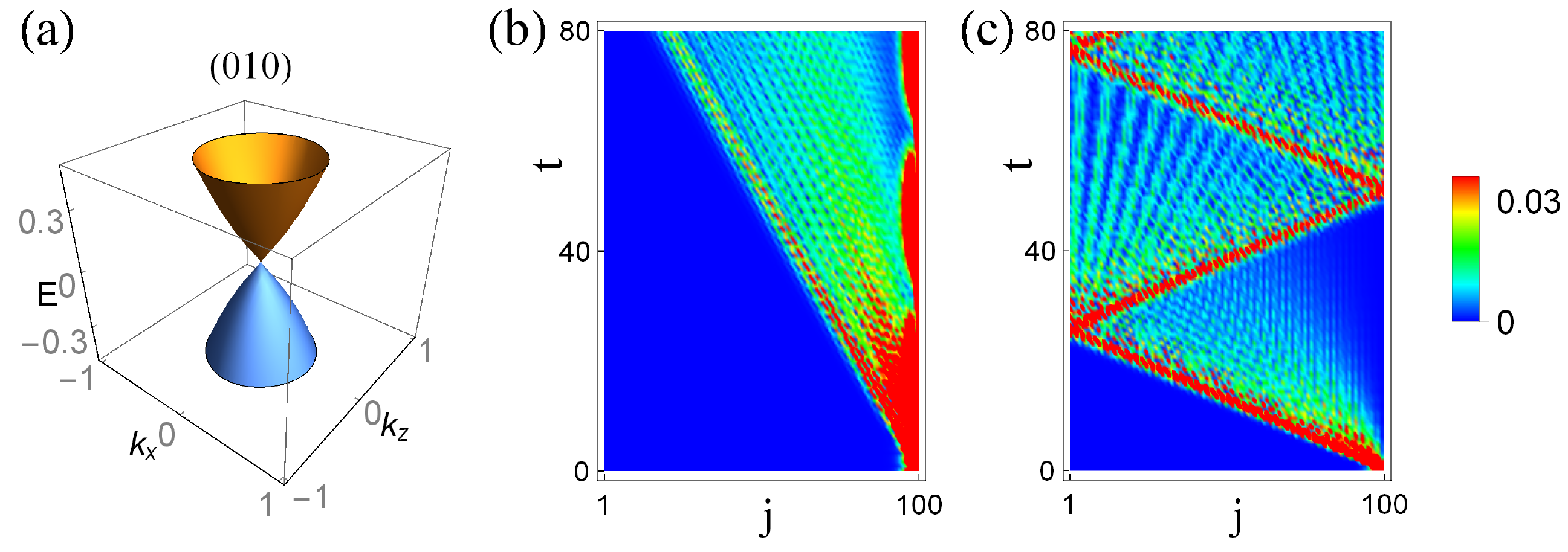}
\caption{Time-evolution of the surface modes of Hopf insulator. (a) Surface states with (010) slab. The Dirac point is located at $(k_x,k_z)=(0,0)$. (b)(c) Time-resolved probability density $\rho(j,t)$ for surface mode at $k_x=k_z=0.1$, $j$ labels the lattice site along $y$ direction. For (b), $M=2.5$, the post-quench Hopf invariant is $\mathcal{L}=1$. For (c), $M=4$, $\mathcal{L}=0$. The pre-quench Hamiltonian is set as $M=1.5$, $\mathcal{L}_0=1$. }
\label{fig5}
\end{figure*}

Similarly, the $\mathbb{Z}_2$ dynamical invariant $\nu$ can be understood as the Jacobian twisting of $\textrm{SO}(3)$ matrices, as sketched in Figs. \ref{fig4}(a)(b). As the base manifold is $\textrm{T}^4$ and the target manifold is $S^3$, the preimage of a point on $S^3$ is a 1D closed loop and the Jacobian is a $3\times 4$ matrix, with three independent column vectors. By traveling along the preimage loop, the twisting of the three orthogonal column vectors (which can be regarded as the three vectors of an $\textrm{SO}(3)$ matrix) gives the desired $\mathbb{Z}_2$ invariant, characterized by 
\begin{eqnarray}
\pi_4(\textrm{SU}(2))=\pi_1(\textrm{SO}(3))=\mathbb{Z}_2.
\end{eqnarray}
The $\mathbb{Z}_2$ nature is followed by the standard belt trick argument \cite{book,belt}. In the above examples [see Sec. \ref{sec:defect}. B] with trivial initial state $|\xi_0\rangle=(1,0)^T$, the $\pi$-defect is nothing but the preimage of $|\xi_0\rangle$ in the (3+1)-D momentum-time space ($t=\frac{\pi}{2}$), and the Weyl cone twisting [see Eq. (\ref{twist})] along this loop defect is an example of the general Jacobian twisting.

\section{Time evolution of surface state}\label{sec:edge}
In this section, we discuss the dynamical behaviors of the surface states of the Hopf insulator. To this end, the pre-quench Hamiltonian is set to be topological with Hopf invariant $\mathcal{L}_0=1$. The quench is implemented by a sudden change of the mass $M$ in model (\ref{model}). For Hopf insulator, depending on the chosen slab geometry, there are two types of surface modes \cite{hopf1}. For (010) or (100) slab, the surface states form a single Dirac point, as shown in Fig. \ref{fig5}(a); while for (001) slab, the surface states form a ring of gapless points. The time evolution of the two types of surface modes exhibit similar behaviors. For simplicity, we focus on the single Dirac point.

We denote $|\psi_{surf}\rangle$ as the pre-quench surface state in the bulk gap. After the quench, $|\psi_{surf}\rangle$ is no longer an eigenstate of the post-quench Hamiltonian $H$ and evolves accordingly. The time-resolved probability density at site $j$ of the time-evolved state is defined as 
\begin{eqnarray}
\rho(j,t)=|\langle j|e^{-i Ht}|\psi_{surf}\rangle|^2.
\end{eqnarray}
We numerically calculate the probability density for two different post-quench Hamiltonians, with $\mathcal{L}=1$ (but a different M value) and $\mathcal{L}=0$, respectively. The bulk dynamical topology for the two cases are characterized by $\nu=0$ and $\nu=1$, respectively. For the former case as shown in Fig. \ref{fig5}(b), the probability density remains spatially localized retaining its initial profile due to large overlapping between surface modes of pre-/post-quench Hamiltonians; For the latter case as shown in Fig. \ref{fig5}(c), the wavetrain moves away from the boundary and enters linearly into the bulk. During time evolution, the density profiles bounce between the two opposite surfaces. For both cases, the spatial profile will decay as a power law at very long time. These results clearly show the two different quench processes indeed have different dynamical behaviors of surface modes. It remains a challenging open problem for future work to relate these dynamical behaviors to the $\mathbb{Z}_2$ topological invariants obtained here.

\section{conclusions and discussions}\label{sec:summary}
In conclusion, we have demonstrated that the quench dynamics of Hopf insulators can be categorized into two topologically distinct classes and characterized by a $\mathbb{Z}_2$ invariant $\nu=(\mathcal{L}-\mathcal{L}_0)\bmod 2$, relating the pre- and post-quench static Hopf invariants. We have constructed this dynamical invariant using the homotopy group $\pi_4(\textrm{SU}(2))=\mathbb{Z}_2$ based on the loop unitary operator. The topological origin of the nontrivial dynamics is further revealed from the emergence of $\pi$-defects in the phase bands. We have provided several perspectives on the $\mathbb{Z}_2$ nature of the quench dynamics.

Our characterization of quench dynamics based on the loop unitary is general and works for other spatial dimensions and other symmetry classes \cite{haiping}. Previously we have utilized this method to characterize the quench dynamics of Chern insulators in 2D, where the loop unitary provides a mapping $\textrm{T}^3\rightarrow S^3$ with homotopy group $\pi_3(\textrm{SU}(2))=\mathbb{Z}$. The reduced $\mathbb{Z}_2$ classification here provides an important example, where the quench dynamics and static band topology possess different types of invariants ($\mathbb{Z}_2$ and $\mathbb{Z}$).


While the $\mathbb Z_2$ invariant may appear abstract, the nontrivial dynamical topology can be extracted by measuring the topological charge of the $\pi$-defect. Recently, two proposals for realizing Hopf insulators in lattice-trapped ultracold fermionic atoms \cite{hopfpro1} (for example $\ce{^6}Li$) and dipolar molecules \cite{hopfpro2} have been put forward. In these proposed systems, the $\pi$-defect and the phase-band winding structures [See Fig. \ref{fig2}] can be directly measured from the time- and momentum-resolved full Bloch-state tomography \cite{quenchexp1,quenchexp2,quenchexp3,azi1,azi2,tomograph1,tomograph2}. 

\section*{Acknowledgments} EZ would like to thank P. Goswami and D. Deng for illuminating discussions on Hopf insulators. This work is supported by AFOSR Grant No. FA9550-16-1-0006 and NSF Grant No. PHY-1707484.

\end{document}